\documentclass[twocolumn,showpacs,preprintnumbers,amsmath,amssymb]{revtex4}
\usepackage{dcolumn}
\usepackage{bm}
\usepackage{graphicx}%
\usepackage{epsfig}
\begin{document}
\title{Anomalous decay of an atom in structured band gap reservoirs}
\author{Filippo Giraldi }
\email{giraldi@ukzn.ac.za, filgi@libero.it} \affiliation{Quantum Research Group,
School of Physics and National Institute for Theoretical Physics,
University of KwaZulu-Natal, Durban 4001, South Africa}
\author{Francesco Petruccione}
\email{petruccione@ukzn.ac.za}
\affiliation{Quantum Research Group, School of Physics and National Institute for Theoretical Physics,
University of KwaZulu-Natal, Durban 4001, South Africa}


\begin{abstract}
We analyze the spontaneous emission of a two-level atom interacting with a special class of structured reservoirs of field modes with band gap edge coinciding with the atomic transition frequency. The exact time evolution of the population of the excited level is evaluated analytically through series of Fox-$H$ functions. Over estimated long time scales, inverse power law relaxations emerge, with powers decreasing continuously to $2$ according to the choice of the special reservoir. No trapping of the population of the excited level emerges. The same results are recovered in presence of $N-1$ atoms, each one in the ground state, described by the Dicke model. The power of the inverse power law decay results to be independent of $N$. A critical number $N_{\alpha}^{\left(\star\right)}$ is evaluated, such that, for $N \gg N_{\alpha}^{\left(\star\right)}$, the inverse power law decay vanishes.
\end{abstract}

\maketitle

\section{Introduction}

The influence of an external environment on the spontaneous emission of an excited atom has been widely studied in literature. Great interest is devoted to the interaction of an atom with a reservoir of field modes, provided by cavities \cite{harochebook,haroche} or dielectric media exhibiting photonic band gap (PBG) \cite{y87,j87,jpc}.  According to the structure of the reservoir, the atomic decay can be either enhanced or suppressed. Important effects emerge if the atomic transition frequency is near the PBG edge.

In Ref. \cite{JQ1994} the decay of an atom placed in a three dimensional periodic dielectric, supplying a PBG \cite{jw0}, is analyzed.
Under certain physical conditions involving the detuning, a portion of the initial population of the excited level is trapped and the dynamics is described by the  photon-atom bond dressed or quasi-dressed states \cite{jw0,jw1}. Over long time scales, the spontaneous emission exhibits an oscillating behavior depending on the detuning between the atomic transition and the band edge frequencies. If the transition frequency lies near (also outside) the band edge frequency, the trapping of the population of the excited level reveals \cite{JQ1994}.

The spontaneous emission of a two-level atom (TLA) has been also studied in presence of $N-1$ unexcited atoms interacting according to the Dicke model \cite{dicke}. The atomic decay of the population of the excited level exhibits a behavior similar to the case of one single atom, with an additional time scale factor equal to $N^{2/3}$ for an isotropic band gap and to $N$ or $N^{2}$ for an anisotropic two dimensional or three dimensional band edges, respectively. For a detailed analysis we refer to \cite{JQ1994}

An attractive model for the inhibition of the atomic emission has been developed in Refs. \cite{lg1} and \cite{lg2}, where the reservoir is dynamically modified through the interaction with an oscillating external bath. Physically, this condition is realized by making the main cavity leak to an external cavity with a moving mirror. The shape of the reservoir is unchanged but the couplings vary on time.

 Structured reservoir described by Lorentzian type spectral densities interacting with a multilevel atom has been analyzed in Ref. \cite{garraway}, \cite{garraway2} and \cite{GD}. The crucial assumption that the frequency range is $\left(-\infty,+\infty\right)$ makes the exact dynamics characterized by the poles of the spectral density in the lower half plane and the corresponding pseudomodes \cite{garrawayknight}.  A proper reservoir described by a spectral density which is the difference of two Lorentzians, exhibits a PBG. If the TLA is resonant with the gap, the population of the excited level is permanently trapped.

Recently \cite{GP,GP2}, an engineering reservoir approach has been adopted in order to delay the decoherence process affecting a qubit due to the interaction with its external environment. Special reservoirs of field modes, interacting with a qubit in rotating wave approximation \cite{JC}, are designed in such a way that the qubit transition frequency coincides with a band gap edge. Over estimated long time scales, the decoherence process results in inverse power law relaxations, with powers decreasing continuously to unity, according to the choice of the special reservoir.

  The first theoretical approach to particle emission was made by Gamow at the beginning of the last Century, interpreting the decay as the appearance of purely outgoing waves in the solution of the corresponding Schr\"odinger equation at large distance and relating the absence of incoming waves to complex energy eigenvalues. Thus, the processes of tunneling and decay consist in resonant states as solutions of the Schr\"odinger equation with purely outgoing boundary conditions. Survival and nonescape probabilities for particle decay exhibit inverse power law behavior with power $3$ over long time scales. A detailed report on the argument and the above results is provided in Ref. {\cite{GarciaCalderon}}.

In this scenario we study the spontaneous emission of an excited TLA interacting with the structured reservoirs introduced in Ref. \cite{GP2}. We aim to describe analytically the exact dynamics of the atomic population of the excited level and compare with the decay provided by a periodic dielectric \cite{JQ1994}. The atomic decay in presence of $N-1$ atoms in the ground state, interacting according to the Dicke model, will also be analyzed.

\section{The model}\label{M}

We consider a TLA interacting with a distribution of field modes  in rotating wave approximation and with dipole interactions \cite{JC, BP,weiss, garraway,garraway2}. By considering a system of units where $\hbar=1$, the
Hamiltonian of the whole system is $H=H_A+H_E+H_I$,
\begin{eqnarray}
&&H_A=\omega_0 |1\rangle_a \,_a \langle 1|,\hspace{2em} H_E=\sum_{k=1}^{\infty} \omega_k \, b^{\dagger}_k b_k,  \nonumber \\ &&H_I=
\imath \sum_{k=1}^{\infty} g_k \left(b^{\dagger}_k\otimes|\,0\rangle_a \,_a\langle 1 |
-b_k \otimes|1\rangle_a \,_a\langle 0 |
 \right).\nonumber
\end{eqnarray}
The state kets $|\,0\rangle_a$ and $|1\rangle_a$ represent the ground and the excited state of the atom, while $b_k^{\dagger}$ and $b_k$ are the creation and annihilation operators, respectively, acting on the Hilbert space of the $k$-th boson and fulfilling the commutation rule
   $\left[b_k,b_{k^{\prime}}^{\dagger}\right]=\delta_{k,k^{\prime}}$ for every
   $k,k^{\prime}=1,2,3,\ldots$. The constants $g_k$, representing the coupling between the atomic transition and the $k$-th field mode, depend on the geometry and are considered to be real without loss of generality.

    Starting from the initial state of the total system
   \begin{equation}
|\Psi(0)\rangle=|1\rangle_a \otimes |0\rangle_E, \label{Psi0}
\end{equation}
   where $|\,0\rangle_E$ is the vacuum state of the environment, the exact time evolution is
   described by the form
   \begin{eqnarray}
   &&|\Psi(t)\rangle=c(t)|1\rangle_a \otimes |0\rangle_E + \sum_{k=1}^{\infty}d_k(t)|\,0\rangle \otimes |k\rangle_E, \nonumber \\
   &&|k\rangle_E=b^{\dagger}_k |0\rangle_E, \hspace{2em}k=1,2,\ldots. \nonumber
\end{eqnarray}
  The dynamics is easily studied in the interaction picture with respect to $H_S+H_E$,
   \begin{eqnarray}
   &&|\Psi(t)\rangle_I=e^{\imath \left(H_S+H_E\right)t}|\Psi(t)\rangle \nonumber \\
&&=C(t)|1\rangle\otimes|0\rangle_E  +\sum_{k=1}^{\infty}\Lambda_k(t)|0\rangle \otimes |k\rangle_E \nonumber,
\end{eqnarray}
where $\imath$ is the imaginary unity, $C(t)=e^{\imath \omega_0 t} \,c(t)$ and $\Lambda_k(t)=e^{\imath \omega_k t}\,d_k(t)$ for every $k=1,2,\ldots$.
The dynamics is described by the system
\begin{eqnarray}
&&\dot{C}(t)=- \sum_{k=1}^{\infty} g_k \, \Lambda_k(t)\,e^{-\imath\left(\omega_k-\omega_0\right)t},  \label{CEq}\\ &&\dot{\Lambda}_k(t)= \,g_k \, C(t)\,e^{\imath\left(\omega_k-\omega_0\right)t}, \hspace{2em} k=1,2,3\ldots, \label{LambdaEq}
\end{eqnarray}
solved in Ref. \cite{JQ1994}. The amplitude $\langle 1| \otimes\, _E\langle 0|| \Psi(t) \rangle_I$, labeled as $C(t)$, is driven by the following convoluted structure equation:
\begin{eqnarray}
\dot{C}(t)=-\left(f\ast C\right)(t), \hspace{2em} C(0)=1,
\label{cMEQ&CorrReservoir}
\end{eqnarray}
 where  $f$ is the two-point correlation function of the reservoir of field modes,
 \begin{eqnarray}
f\left(t-t^{\prime}\right)=\sum_{k=1}^{\infty}
g_k^2 \,e^{-\imath \left(\omega_k-\omega_0\right)
\left(t-t^{\prime}\right)}.
\nonumber
\end{eqnarray}
  For a continuous distribution of modes described by  $\eta\left(\omega\right)$, the correlation function is expressed through the spectral density function $J\left(\omega\right)$,
 \begin{eqnarray}
f\left(\tau\right)=
\int_0^{\infty}J\left(\omega\right) e^{-\imath\left(\omega-\omega_0\right)\tau }
d \omega, \nonumber
\end{eqnarray}
where $J\left(\omega\right)=\eta\left(\omega\right) g\left(\omega\right)^2$ and $g\left(\omega\right)$ is the frequency dependent coupling constant.
The above model was studied by Garraway \cite{garraway,garraway2} and the corresponding exact dynamics was analytically described for Lorentzian type distributions of field modes.

\section{The exact decay}\label{ed}

We analyze the exact time evolution of the population of the excited atomic level in case the structured reservoir is described by the following class of continuous spectral densities:
\begin{eqnarray}
&&J_{\alpha}\left(\omega\right)=
 \frac{2 A \left(\omega-\omega_0\right)^{\alpha}
 \Theta\left(\left(\omega-\omega_0\right)/\omega_0\right)}
 {a^2+\left(\omega-\omega_0\right)^2 },\label{Jalpha} \\ &&A/\, a^{3-\alpha}>0, \hspace{2em}1>\alpha>0. \nonumber
\end{eqnarray}
The special spectral densities exhibit a PBG edge coinciding with the atomic transition frequency, an absolute maximum $M_{\alpha}$ at the frequency $\Omega_{\alpha}$,
 \begin{eqnarray}
 &&M_{\alpha}=J_{\alpha}\left(\Omega_{\alpha}\right)=A\, \alpha^{\alpha/2} a^{\alpha-2}\left(2-\alpha\right)^{1-\alpha/2},  \nonumber \\ &&\Omega_{\alpha}=\omega_0+a \,\alpha^{1/2}\left(2-\alpha\right)^{1/2},\nonumber
 \end{eqnarray}
and are piece-wise similar to those usually adopted, i.e. sub-ohmic at low frequencies, $\omega/\,\omega_0 \gtrsim 1$, and inverse power law at high frequencies, $\omega /\,\omega_0\gg 1$, similar to the Lorentzian one, though with different power,
\begin{eqnarray}
&&J_{\alpha}\left(\omega\right)\sim
 2 A/a^2 \left(\omega-\omega_0\right)^{\alpha}, \hspace{2em}  \omega \to \omega_0^+, \nonumber
 \\\hspace{1em} &&J_{\alpha}\left(\omega\right)\sim
 2 A \,\omega^{\alpha-2}, \hspace{2em} \omega \to +\infty. \nonumber
 \end{eqnarray}

 The convoluted structure equation (\ref{cMEQ&CorrReservoir})
 has been studied in detail in Ref. \cite{GP2} for correlation functions $f$, generated by the spectral densities (\ref{Jalpha}). The particular case $\alpha=1/2$, characterized by a simple dynamics, has been separately analyzed in Ref. \cite{GP}. For the sake of clarity we report the main results and refer to the mentioned literature for a detailed proof.

 \subsection{The particular case $\alpha=1/2$}

 The dynamics described by Eq. (\ref{cMEQ&CorrReservoir}), takes a simple form in the special case $\alpha=1/2$. The solution is
 \begin{equation}
 C(t)=\frac{1}{ \sqrt{\pi}}\sum_{l=1}^4
  \chi_l \,R\left(\chi_l\right)\,
 e^{ \chi_l^2 t}\,\Gamma\left(1/2,\chi_l^2 t\right),\label{Gt}
\end{equation}
  where $R(z)$ is a rational function,
  \begin{equation}
   R(z)=\frac{\left(1-\imath\right)
   \left(a^{1/2}+z\right)\left(\imath \,a^{1/2}+ z\right)}
   { 2 z \left(\left(1+\imath\right)a+3 \,a^{1/2} z +2 \left(1-\imath\right)z^2\right) },
   \label{R}
   \end{equation}
  while the complex numbers $\chi_1,\chi_2,\chi_3$ and $\chi_4$, are the roots, \emph{distinct} for every positive value of both $A$ and $a$, of the polynomial
   \begin{equation}
    Q(z)=\pi \sqrt{2/a} \,A +\imath \,a \,z^2+ \left(1+\imath\right)a^{1/2} z^{3}+z^4. \label{Q}
   \end{equation}
 The analytical expressions of the roots are given by Eqs. (\ref{chi1}), (\ref{chi2}), (\ref{chi3}) and (\ref{chi4}) of Appendix \ref{A}.
 Finally, the time evolution of the population of the excited state, $P(t)$, given by the term $\left|\,C(t)\right|^2$, reads
 \begin{eqnarray}
 P(t)=\frac{1}{ \pi}\left|\,\sum_{l=1}^4
  \chi_l\,R\left(\chi_l\right)\,
 e^{ \chi_l^2 t}\,\Gamma\left(1/2,\chi_l^2 t\right)\right|^{\,2}. \label{P}
 \end{eqnarray}
  The decay, fulfilling the initial condition $P(0)=1$, is described by a simple quadratic form of Incomplete Gamma functions.

\subsection{The case $1>\alpha>0$}

 In case the reservoir of field modes is described by the class of spectral density (\ref{Jalpha}), the population of the excited level is the square modulus of the function $C_{\alpha}(t)$, solution of Eq. (\ref{cMEQ&CorrReservoir}), where $f$ is the corresponding correlation function of the reservoir. Following the arguments of Ref. \cite{GP2}, the searched solution,
 \begin{eqnarray}
    &&C_{\alpha}(t)=\sum_{n=0}^{\infty}\sum_{k=0}^n\frac{(-1)^n\, z_{\alpha}^k \,z_0^{n-k}\,t^{3n-\alpha k}}{k!(n-k)!} \nonumber \\  &&\times\Bigg(H_{1,2}^{1,1}\left[z_1 t^2\Bigg|
      \begin{array}{rr}
\left(-n,1\right) \hspace{5em}
\\
\left(0,1\right), \left(\alpha k-3 n,2\right)
\end{array}
\right]
     \nonumber \\ &&-\,a^2 t^2 H_{1,2}^{1,1}\left[z_1 t^2\Bigg| \begin{array}{rr}
\left(-n,1\right)\hspace{6.7em}
\\
\left(0,1\right), \left(\alpha k-3 n-2,2\right)
\end{array}
\right]\Bigg)
     ,\,\, \label{GaH}
    \end{eqnarray}
 results in a series of Fox $H$-functions, defined through a Mellin-Barnes type integral in the complex domain,
  \begin{eqnarray}
&&H_{p,q}^{m,n}\left[z\Bigg|
\begin{array}{rr}
\left(a_1,\alpha_1\right), \ldots,\left(a_p,\alpha_p\right)\\
\left(b_1,\beta_1\right), \ldots,\left(b_q,\beta_q\right)\,\,
\end{array}
\right]
=\frac{1}{2 \pi \imath}\nonumber \\&& \times \int_{\mathcal{C}}
\frac{\Pi_{j=1}^m \Gamma\left(b_j+\beta_j s\right)
\Pi_{m=1}^n \Gamma\left(1-a_l-\alpha_l s\right)
z^{-s}}{\Pi_{l=n+1}^p \Gamma\left(a_l+\alpha_l s\right)
\Pi_{j=m+1}^q \Gamma\left(1-b_j-\beta_j s\right)
} \,  ds,\nonumber
\end{eqnarray}
under the conditions that the poles of the Gamma functions in the dominator, do not coincide. The empty products are interpreted as unity. The natural numbers $m,n,p,q$ fulfill the constraints: $0\leq m\leq q$, $0\leq n\leq p$, and $\alpha_i,\beta_j\in \left(0,+\infty\right)$ for every $i=1,\cdots,p$ and $j=1,\cdots,q$. For the sake of shortness, we refer to \cite{HbookMSH} for details on the contour path $\mathcal{C}$, the existence and the properties of the Fox $H$-function. The Generalized Mittag-Leffler  \cite{Prabhakar,MainardiBook}, the Generalized Hypergeometric, the Wright \cite{WrightKST} and the Meijer $G$-functions \cite{em2} are particular cases of the Fox $H$-function, thus, the amplitude $C_{\alpha}(t)$ can be expressed as a series of each Special function mentioned above.
   The parameters involved in Eq. (\ref{GaH}) are defined as follows:
    \begin{eqnarray}
    &&z_1=\pi A a^{\alpha-1}\sec\left(\pi \alpha /2\right)-a^2, \hspace{1em} z_0=\imath \pi A a^{\alpha} \cos\left(\pi \alpha/2\right), \nonumber \\ &&z_{\alpha}=-2\imath \pi A e^{-\imath \pi \alpha/2} \csc\left(\pi \alpha\right).\nonumber
\end{eqnarray}

  Simplified forms are obtained in particular cases. For example, the condition $A=A^{\left(\star\right)}$,
where $A^{\left(\star\right)}= a^{3-\alpha}\, \cos\left(\pi \alpha/2\right)/\pi$,
 corresponding to $z_1=0$, gives a power series solution,
 \begin{eqnarray}
         &&C_{\alpha}^{\left(\star\right)}(t)
 = \sum_{n=0}^{\infty}\sum_{k=0}^n\frac{(-1)^n\, n!\, z_{\alpha}^k \,z_0^{n-k}\,t^{3n-\alpha k}}{k!\,(n-k)!\,\Gamma\left(3n-\alpha k+1\right)}
    \Bigg\{ 1\nonumber \\ &&-\,a^2\,\frac{ \,\Gamma\left(3n-\alpha k+1\right)\, }{\Gamma\left(3n-\alpha k+3\right)}\,t^2 \Bigg\}, \hspace{2em} 1>\alpha>0. \label{Ghyperz10}
        \end{eqnarray}
In case the parameter $\alpha$ takes rational values, $p/q$, where $p$ and $q$ are distinct prime numbers such that $0<p<q$, the solution of Eq. (\ref{cMEQ&CorrReservoir}) can be expressed as a modulation of exponential relaxations \cite{NM},
\begin{equation}
 C_{p/q}(t)=\int_0^{\infty}d\eta \,\int_0^{\infty}d\xi\, \Phi_{p/q}\left(\eta, \xi\right) e^{-\xi t},
 \label{GpqtI}
 \end{equation}
 where
 \begin{eqnarray}
 &&\Phi_{p/q}\left(\eta, \xi\right)=\sum_{l=1}^n\sum_{k=1}^{m_l} \frac{b_{l,\,k}\left(\zeta_l\right)}{\pi}\,
 \, \eta^{m_l-k} \nonumber \\&& \times \sin\left(\eta\, \xi^{1/q}\sin \left(\pi/q\right)\right)e^{\eta\left(\zeta_l- \cos\left(\pi/q\right)\xi^{1/q}\right)}. \nonumber
  \end{eqnarray}
   The rational functions $b_{l,\,k}\left(z\right)$ read
  \begin{eqnarray}
  b_{l,\,k}\left(z\right)=\frac{d^{k-1}/dz^{k-1}\,\left[\left(z^q-a\right)
  \left(z^q+a\right)\left(z-\zeta_l\right)^{m_l}/ Q\left(z\right)\right]}
  { \left(m_l-k\right)!\left(k-1\right)!}, \nonumber
   \end{eqnarray}
   for every $l=1,\ldots,n$, and $k=1,\ldots ,m_l$. The complex numbers $\zeta_1, \ldots, \zeta_n$ are the roots of the polynomial
 \begin{equation}
 Q(z)=z^{3q}+z_1\, z^q+z_{\alpha}\,z^{p}+z_0 \label{Q}
 \end{equation}
  and $m_l$ is the multiplicity of $\zeta_l$, for every $l=1,\ldots,n$, which means $Q(z)=\Pi_{l=1}^n \left(z-\zeta_l\right)^{m_l}$ and $\sum_{l=1}^{n}m_l=3\, q$.

   The particular conditions $\alpha=3/4$ and $A=a^{9/4}\cos\left(3 \pi/8\right)/ \pi$ give a vanishing value of $z_1$ and the roots $\zeta_1,\ldots,\zeta_l$ can be evaluated analytically from the solutions of a quartic equation. For the sake of shortness, the corresponding expressions are omitted. The remaining cases of rational values of $\alpha$, force to a numerical evaluation of the roots of $Q(z)$, once the numerical values of both $a$ and $A$ are fixed.
  The necessary analysis of the function $C_{\alpha}(t)$ is now completed.

  Starting from the initial condition (\ref{Psi0}), the interaction with a a reservoir of field modes described by the spectral density $J_{\alpha}\left(\omega\right)$, gives the following time evolution of the population of the excited level:
  \begin{eqnarray}
     &&P_{\alpha}(t)=\Bigg|\sum_{n=0}^{\infty}\sum_{k=0}^n\frac{(-1)^n\, z_{\alpha}^k \,z_0^{n-k}\,t^{3n-\alpha k}}{k!(n-k)!}\nonumber \\ &&\times
      \Bigg(H_{1,2}^{1,1}\left[z_1 t^2\Bigg|
      \begin{array}{rr}
\left(-n,1\right) \,\,\,\,\,\,\,\,\,\,\,\,\,\,\,\,\,
\,\,\,\,\,\,\,\,\,\,\,\,\,\,
\\
\left(0,1\right), \left(\alpha k-3 n,2\right)
\end{array}
\right] -\,a^2 t^2\nonumber \\
     && \times\, H_{1,2}^{1,1}\left[z_1 t^2\Bigg| \begin{array}{rr}
\left(-n,1\right)\,\,\,\,\,\,\,\,\,\,\,\,\,\,\,\,\,
\,\,\,\,\,\,\,\,\,\,\,\,\,\,\,\,\,\,\,\,\,\,\,\,\,
\\
\left(0,1\right), \left(\alpha k-3 n-2,2\right)
\end{array}
\right]\Bigg)\Bigg|^{\,2}
     . \label{Pa}
         \end{eqnarray}
  The decay fulfills the initial condition $P_{\alpha}(0)=1$ and gets a simplified form in the particular cases reported above.

\section{Inverse power laws}\label{ipl}
The theoretical analysis of the \emph{exact} decay, performed in the Section \ref{ed}, leads to concrete results. The special reservoir described by the spectral density $J_{1/2}\left(\omega\right)$ gives an exact decay described by a quadratic form of Incomplete Gamma functions and the corresponding asymptotic expansions \cite{em2} identify a \emph{time scale} $\tau$ and a \emph{decay factor} $\zeta$,
  \begin{eqnarray}
 &&\tau=\max \left\{\left|\chi_l\right|^{-2},\,l=1,2,3,4\right\},
  \label{tauE}\\ && \zeta=\frac{1}{4 \pi }
  \Big|\sum_{l=1}^4  R\left(\chi_l\right)\,
 \chi_l^{-2}\Big|^{\,2}, \nonumber
 \end{eqnarray}
 such that, over long time scales, $t/\,\tau\gg1$, the population of the excited level
decays according to the following asymptotic form:
   \begin{eqnarray}
 P(t)\sim \zeta \,t^{-3}
  , \hspace{2em} t\to+\infty. \label{Gasympt}
 \end{eqnarray}

The reservoirs described by the spectral densities $J_{\alpha}\left(\omega\right)$, for every $\alpha \in (0,1)$, provoke an atomic decay resulting in inverse power laws over evaluated long \emph{time scales} $\tau_{\alpha}$.
For $t/\, \tau_{\alpha}\gg 1$ the population of the excited level, $P_{\alpha}(t)$ is described by the asymptotic form
\begin{equation}
P_{\alpha}(t)\sim
   \zeta_{\alpha} \, t^{-2\left(1+ \alpha\right)},\hspace{1em} t\to +\infty,\hspace{2em}1>\alpha>0,\label{Gasympt}
   \end{equation}
   where
   \[\zeta_{\alpha}= \frac{4 \,\alpha^2\, a^{4\left(1-\alpha\right)}\csc^2\left(\pi \alpha\right)\sec^4\left(\pi \alpha/2\right)}{\pi^2 A^2\,
    \left(\Gamma\left(1-\alpha\right)\right)^2}.
   \]
   A simple choice of the time scale is
    \begin{equation}
\tau_{\alpha}=\max\left\{1,\left|\frac{3}{z_0}\right|^{1/3},
\left|\,3\,\frac{z_{\alpha}}{z_0}\right|^{1/\alpha},
3\left|\frac{z_1}{z_0}\right| \right\}.
 \label{taua}
\end{equation}
Notice that the choice of the time scale for inverse power laws is not unique: in general, the time scales $\tau$ and $\tau_{1/2}$ are different.

\begin{figure}[t]
\centering
\includegraphics[height=4.25 cm, width=8.25 cm]{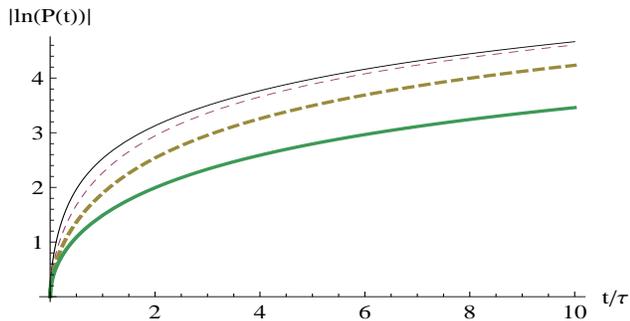}
\vspace{0.2cm}
\caption{The time evolution of $\left| \ln \left(P(t)\right)\right|$, the absolute value of the natural logarithm of population of the excited level obtained from Eq. (\ref{P}), of a TLA interacting with the distribution of field modes $J_{1/2}\left(\omega\right)$, given by Eq. (\ref{Jalpha}) for $\alpha=1/2$, for $0\leq t/\tau\leq 10$. The curve $\gamma_1$ (thick solid line) is obtained for $A=(a/1000)^{5/2}$, $\gamma_2$ (thick dashed line) for $A=100 \,a^{5/2}$, $\gamma_3$ (thin dashed line) for $A=a^{5/2}$, $\gamma_4$ (thin solid line) for $A=(a/2)^{5/2} $. The time scales for inverse power laws, given by Eq. (\ref{tauE}), read $\tau\simeq 1.40 /a $ for $\gamma_4$, $\tau \simeq 0.84/a$ for $\gamma_3$, $\tau\simeq 0.06/a$ for $\gamma_2$, $\tau\simeq 7121.40/a$ for $\gamma_1$, respectively.}
\label{graphP12}
\end{figure}

   The results obtained for the case $\alpha=1/2$ are confirmed by the numerical data plotted of Fig. ~\ref{graphP12}.
   Over long time scales, the population of the excited atomic energy level decays with an inverse power law, of power decreasing continuously to $2$, according to the choice of the special reservoir.

\section{N atoms }\label{N}

We extend the study of the spontaneous emission of an excited TLA in presence of $N-1$ ones in the ground state described by the Dicke model \cite{dicke,harochebook} and interacting with the continuous distributions of modes (\ref{Jalpha}).
Following Ref. \cite{JQ1994}, the Hamiltonian of the whole system reads
\begin{eqnarray}
&&H_N=
\sum_{k=1}^{\infty}\left(\omega_k-\omega_0\right) b^{\dagger}_k b_k +
\imath \sum_{k=1}^{\infty} g_k\left(J_{1,\,0} \,b^{\dagger}_k-J_{0,1}\, b_k\right), \nonumber \\
&&J_{l,\,m}=\sum_{n=1}^N |\,l\rangle_{(n)}\,_{(n)}\langle m|,\hspace{2em} l,m=0,1,
\nonumber
\end{eqnarray}
 where $|\,1\rangle_{(n)}$ and $|\,0\rangle_{(n)}$ are the excited and ground state of the $n$-th atom, respectively. Notice that $\hbar=1$ and that the couplings between each atomic transition and the field modes are purely imaginary, i.e., the constants $g_k$ are real for every $k=1,2,\ldots$.

We adopt the following notation: the kets $|J,M\rangle$ are the normalized eigenstates of the operator $J_3=\left(J_{2,\,2}-J_{1,1}\right)/2$, corresponding to the quantum number $M$, and the operator $J^2=J_3^2+\left(J_{2,1} J_{1,\,2}+J_{1,\,2} \,J_{2,1}\right)/2$, corresponding to the quantum number $J$.
The $N$ atoms are initially in the superradiant states $|J,M=1-J\,\rangle$ with only one atom versing in the excited states and the reservoir, unentangled from the atoms, is in the vacuum states. The time evolution is described by the forms:
\begin{eqnarray}
&&\Psi_N(t)\rangle_I=C_N(t)\,|J,M=1-J\,\rangle\otimes |0\rangle_E \nonumber \\
&&+\sum_{k=1}^{\infty}\Lambda_{N,\,k}(t)\, e^{-\imath  \left(\omega_k-\omega_0\right)t}\,|J,M=-J\,\rangle\otimes |k\rangle_E, \nonumber \\
&&
\dot{C}_{N}(t)=-\sqrt{N}\,\sum_{k=1}^{\infty} g_k \,\Lambda_{N,\,k}(t)\,e^{-\imath \left(\omega_k-\omega_0\right)t}, \nonumber \\ &&\dot{\Lambda}_{N,\,k}(t)=\sqrt{N}\, g_k \,C_N(t)\, e^{\imath \left(\omega_k-\omega_0\right)t}, \hspace{2em} k=1,2,3,\ldots. \nonumber
\end{eqnarray}
The above system is solved in Ref. \cite{JQ1994} and leads to the following convoluted structure equation:
\begin{eqnarray}
\dot{C}_N(t)=-\left(C_N\ast f_N\right)(t),\hspace{2em} C_N(0)=1,
\label{CN}
\end{eqnarray}
where $f_N$ is the two-points correlation function of the reservoir,
\begin{eqnarray}
f_N\left(t-t^{\prime}\right)=N \sum_{k=1}^{\infty}g_k^2\, e^{-\imath \left(\omega_k-\omega_0\right)\left(t-t^{\prime}\right)}.
\nonumber
\end{eqnarray}
 For a continuous distribution of field modes, the correlation function is expressed through the spectral density $J\left(\omega\right)$,
\begin{eqnarray}
f_N\left(\tau\right)=N \int_0^{\infty} J\left(\omega\right) e^{-\imath \left(\omega-\omega_0\right)\tau} d\omega. \nonumber
\end{eqnarray}
The corresponding dynamics is obtained by replacing $A$ with $N\, A$ in the corresponding expressions obtained in Sections \ref{ed} and \ref{ipl}.

For the special case $\alpha=1/2$, the time evolution of the population of the excited level is described through the roots $\chi_{N,1}$, $\chi_{N,2}$, $\chi_{N,3}$ and $\chi_{N,4}$ of the polynomial
\begin{equation}
    Q_N(z)=\pi \sqrt{2/a} \,N A +\imath \,a \,z^2+ \left(1+\imath\right)a^{1/2} z^{3}+z^4, \label{QN}
   \end{equation}
obtained by Eqs. (\ref{chi1}), (\ref{chi2}), (\ref{chi3}) and (\ref{chi4}) of Appendix \ref{A} by replacing the parameter $A$ with $A\,N $. The population of the excited state results to be
\begin{eqnarray}
 &&P_N(t)=\frac{1}{ \pi}\Big|\,\sum_{j=1}^4
  \chi_{N,\,j}\,R\left(\chi_{N,\,j}\right)\,
 e^{ \left(\chi_{N,\,j}\right)^2 t} \nonumber \\ &&\times \, \Gamma\left(1/2,\left(\chi_{N,\,j}\right)^2 t\right)\Big|^{2} \label{PN12}
 \end{eqnarray}
and fulfills the initial condition $P_N(0)=1$.

The general case where $1>\alpha>0$ is treated by replacing the parameters $z_0,z_1$ and $z_{\alpha}$ with $z_{N,\,0}, z_{N,\,1}$ and $z_{N,\,\alpha}$, respectively, defined by changing $A$ with $A \,N$,
\begin{eqnarray}
    &&z_{N,\,1}=\pi A N a^{\alpha-1}\sec\left(\pi \alpha /2\right)-a^2, \hspace{1em}z_{N,\,0}=N z_0,\nonumber \\ &&z_{N,\,\alpha}=N z_{\alpha}.\nonumber
\end{eqnarray}
The time evolution of the population of the excited level, $P_{N,\, \alpha}(t)=\left|\,C_{N,\,\alpha}(t)\right|^2$, reads
\begin{eqnarray}
    &&P_{N,\, \alpha}(t)=\Bigg|\sum_{n=0}^{\infty}\sum_{k=0}^n\frac{(-N)^n\, z_{\alpha}^k \,z_{0}^{n-k}\,t^{3n-\alpha k}}{k!(n-k)!} \nonumber \\ &&\Bigg(H_{1,\,2}^{1,1}\left[z_{N,1} \, t^2\Bigg|
      \begin{array}{rr}
\left(-n,1\right) \,\,\,\,\,\,\,\,\,\,\,\,\,\,\,\,\,
\,\,\,\,\,\,\,\,\,\,\,\,\,\,
\\
\left(0,1\right), \left(\alpha k-3 n,2\right)
\end{array}
\right] -\,a^2 t^2\nonumber \\
     && \times H_{1,\,2}^{1,1}\left[z_{N,1}\, t^2\Bigg| \begin{array}{rr}
\left(-n,1\right)\,\,\,\,\,\,\,\,\,\,\,\,\,\,\,\,\,
\,\,\,\,\,\,\,\,\,\,\,\,\,\,\,\,\,\,\,\,\,\,\,\,
\\
\left(0,1\right), \left(\alpha k-3 n-2,2\right)\end{array}
\right]\Bigg)\Bigg|^{\,2}
      \,\, \label{PNa}
   \end{eqnarray}
and fulfills the initial condition $P_{N,\, \alpha}(0)=1$.

   Simplified forms are obtained in particular cases. For example, the condition $A=A_N^{\left(\star\right)}$, defined as follows: $
A_N^{\left(\star\right)}= a^{3-\alpha}{\pi}\, N\, \cos\left(\pi \alpha/2\right)/ \pi$,
 corresponding to $z_{N,1}=0$, gives the following power series solution,
 \begin{eqnarray}
         && C_{N,\,\alpha}^{\left(\star\right)}(t)
 = \sum_{n=0}^{\infty}\sum_{k=0}^n\frac{(-N)^n\, n!\, z_{\alpha}^k \,z_{0}^{n-k}\,t^{3n-\alpha k}}{k!\,(n-k)!\,\Gamma\left(3n-\alpha k+1\right)}
  \nonumber \\ &&\times \Bigg\{ 1-a^2\,\frac{ \,\Gamma\left(3n-\alpha k+1\right)\, }{\Gamma\left(3n-\alpha k+3\right)}\,t^2 \Bigg\}, \hspace{1em} 1>\alpha>0.   \label{GNhyperz10}
        \end{eqnarray}

As regards the dynamics over long time scales, it is described by inverse power laws as well as the case of one atom. For $\alpha=1/2$, a time scale $\tau_N$ and a decay factor $\zeta_N$ emerge,
\begin{eqnarray}
 \tau_N=\max \left\{\left|\chi_{N,\,l}\right|^{-2},\,l=1,2,3,4\right\},
 \label{tauNE}
 \end{eqnarray}
 such that, over long time scales, $t\gg\tau_N$, the population of the excited level
 is described by the asymptotic form
   \begin{eqnarray}
 &&P_N(t)\sim \zeta_N \,t^{-3}
  , \hspace{2em} t\to+\infty, \nonumber \\
  &&\zeta_N=\frac{1}{4 \pi }
  \Big|\sum_{l=1}^4  R\left(\chi_{N,\,\,l}\right)\,
 \left(\chi_{N,\,l}\right)^{-2}\Big|^{\,2}\label{GNasympt}.
 \end{eqnarray}

As regards the general case, $1>\alpha>0$, a \emph{time scale} reveals,
\begin{equation}
\tau_{N,\,\alpha}=\max\left\{1,\left|\frac{3}{z_{N,\,0}}\right|^{1/3},
\left|\,3\,\frac{z_{\alpha}}{z_{0}}\right|^{1/\alpha},
3\left|\frac{z_{N,\,1}}{z_{N,\,0}}\right| \right\},
 \label{tauNa}
\end{equation}
 such that, for $t/\,\tau_{N,\,\alpha}\gg 1 $,
the population of the excited level exhibits
inverse power law behavior described by the asymptotic form
\begin{eqnarray}
&&P_{N,\, \alpha}(t)\sim
   \zeta_{N,\,\alpha} \, t^{-2\left(1+ \alpha\right)},\hspace{2em} t\to +\infty, \label{GasymptN}
      \\ &&\zeta_{N,\,\alpha}= \frac{4 \,\alpha^2\, a^{4\left(1-\alpha\right)}\csc^2\left(\pi \alpha\right)\sec^4\left(\pi \alpha/2\right)}{\pi^2 A^2 N^2\, \left(\Gamma\left(1-\alpha\right)\right)^2},\hspace{0.5em} 1>\alpha>0. \nonumber
   \end{eqnarray}
       Notice that for $N\to +\infty$,
       the time scale for inverse power law tends to the asymptotic value $
       \tau_{\infty,\,\alpha}=\max\left\{1,
\left|\,3/\left(\sin\left(\pi \alpha/2\right)
\cos^2\left(\pi \alpha/2\right)\right)\right|^{\,1/\alpha} /a\right\}$,
 while the decoherence factor vanishes, $\zeta_{N,\,\alpha}\to 0$, which mean that the inverse power law behavior vanishes as the number of atoms increase. We
 define a natural number $N_{\alpha}^{\left(\star\right)}$ such that $\zeta_{N_{\alpha}^{\left(\star\right)},\, \alpha}\,a^{2\left(1+\alpha\right)}\simeq1$,
 \begin{equation}
 N_{\alpha}^{\left(\star\right)}=\left[ \frac{2 \,\alpha\, a^{3-\alpha}\csc\left(\pi \alpha\right)\sec^2\left(\pi \alpha/2\right)}{\pi A \, \Gamma\left(1-\alpha\right)}\right], \hspace{2em} 1>\alpha>0, \label{Nastar}
 \end{equation}
where the square brackets indicate the integer part.
The natural number $N_{\alpha}^{\left(\star\right)}$ estimates the critical order of magnitude for the number of atoms such that inverse power laws appear if the order of magnitude of $N$ is less or equal to $N_{\alpha}^{\left(\star\right)}$ and vanishes if $N\gg N_{\alpha}^{\left(\star\right)}$. For every fixed value of the parameter $\alpha$, the critical value can be enlarged by increasing the ratio $a^{3-\alpha}/ A$.

\begin{figure}[t]
\centering
\includegraphics[height=4.25 cm, width=8.25 cm]{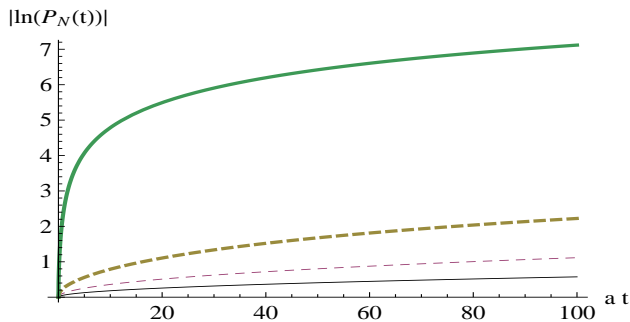}
\vspace{0.2cm}
\caption{The time evolution absolute value of the natural logarithm of $P_{N}(t)$, given by Eq. (\ref{PN12}), the population  of the excited energy level of a TLA interacting with a continuous distribution of field modes described by $J_{1/2}\left(\omega\right)$, given by Eq. (\ref{Jalpha}) for $\alpha=1/2$, in presence of $N-1$ atoms in the ground level, described by the Dicke model, for $0\leq at \leq 100$ and $A=10^{-4}a^{5/2}$. The curve $C_1$ (thin solid line) corresponds to $N=3$, $C_2$ (thin dashed line) to $N=12$, $C_3$ (thick dashed line) to $N=60$ and $C_4$ (thick solid line) to $N=500$. The corresponding time scales for inverse power laws, given by Eq. (\ref{tauNa}) for $\alpha=1/2$, read $\tau_{10000,1/2}\simeq 0.84/a$, $\tau_{60,1/2}\simeq46.1/a$, $\tau_{12,1/2}\simeq 206.9/a$, $\tau_{3,1/2}\simeq 789.0/a$. The critical number, given by Eq. (\ref{Nastar}) for $\alpha=1/2$, is $N^{\left(\star\right)}_{1/2}=3591$.}
\label{graphPN12}
\end{figure}

Again, the result obtained for the case $\alpha=1/2$, are confirmed by the numerical data plotted of Fig. ~\ref{graphPN12}.
   Over long time scales, the population of the excited atomic energy level decays with an inverse power law, of power decreasing continuously to $2$, according to the choice of the special reservoir.


\section{Conclusions}

We study the spontaneous emission of a TLA interacting with a class of structured reservoirs exhibiting a PBG edge coinciding with the atomic transition frequency, and piece-wise similar to those usually adopted, i.e., sub-ohmic at low frequencies and inverse power laws at high frequencies, similar to the Lorentzian one.
The exact time evolution of the population of the excited level is evaluated analytically through series of Fox-H functions.
Over evaluated long time scales the population of the excited level vanishes according to inverse power laws, with powers decreasing continuously to $2$ depending on the choice of the special reservoir.

    We also consider also the decay of an excited atom in presence of $N-1$ atoms, each one in the ground state, described by the Dicke model, interacting with the special class of reservoirs. Again, the exact time evolution of the population of the excited level is described analytically through series of Fox-$H$ functions and inverse power laws emerge over estimated long time scales.  Though the powers are not affected by $N$, the inverse power law decay tends to vanish if $N\gg N_{\alpha}^{\left(\star\right)}$, where $N_{\alpha}^{\left(\star\right)}$ is a properly defined critical number. No trapping of the population reveals.

The above results are of interest if compared to the following ones.
An atom interacting with a radiation field created by a three-dimensional periodic dielectric with transition frequency lying near the edge of the PBG, exhibits an oscillatory decay of the population of the excited level with trapping, spectral splitting and subnatural linewidth. In case the atom is surrounded by $N-1$ atoms in ground states, the time scale factor of the decay is $N^{2/3}$ for an isotropic band gap, $N$ or $N^2$ for anisotropic two-dimensional or three dimensional edges, respectively \cite{JQ1994}.

The special class of reservoirs of modes can in principle be realized with materials providing the PBG structure. A $N$-period one dimensional lattice provides a band gap through an appropriate sequence of dielectric unit cells \cite{bendickson}. The corresponding density of frequency modes depends on the transmission coefficients of each unit cell. Also, the advanced technologies concerning diffractive grating and photonic crystals allows the realization of tunable 1D PBG microcavities \cite{PBGMC1,PBGMC2,PBGQED}. The action of such a structured environment on a qubit could be a way of delaying the decoherence process with fundamental applications to Quantum Information processing Technologies.

\appendix
\section{details}\label{A}
The exact expressions of the roots of the fourth order polynomial $Q(z)$, given by Eq. (\ref{Q}), are described by the following forms:
\begin{widetext}
\begin{eqnarray}
&&\chi_1=-\frac{\left(1+\imath\right) a^{1/2}}{4}+\frac{\gamma}{2}+\frac{1}{2}
\Bigg(\frac{\left(\imath-1\right)a^{3/2}}{2\, \gamma}+ \frac{a^2-12 \pi\sqrt{2/a}A}{3 \,\phi}-\imath \frac{a}{3}-\frac{\phi}{3}\Bigg)^{1/2}, \label{chi1} \\
&&\chi_2=-\frac{\left(1+\imath\right) a^{1/2}}{4}+\frac{\gamma}{2}-\frac{1}{2}
\Bigg(\frac{\left(\imath-1\right)a^{3/2}}{2 \, \gamma}+ \frac{a^2-12 \pi\sqrt{2/a}A}{3\, \phi}-\imath \frac{a}{3}-\frac{\phi}{3}\Bigg)^{1/2}, \label{chi2} \\
&&\chi_3=-\frac{\left(1+\imath\right) a^{1/2}}{4}-\frac{\gamma}{2}+\frac{1}{2}
\Bigg(\frac{\left(1-\imath\right)a^{3/2}}{2\, \gamma}+ \frac{a^2-12 \pi\sqrt{2/a}A}{3\, \phi}-\imath \frac{a}{3}-\frac{\phi}{3}\Bigg)^{1/2}, \label{chi3} \\
&&\chi_4=-\frac{\left(1+\imath\right) a^{1/2}}{4}-\frac{\gamma}{2}-\frac{1}{2}
\Bigg(\frac{\left(1-\imath\right)a^{3/2}}{2\, \gamma}+ \frac{a^2-12 \pi\sqrt{2/a}A}{3 \, \phi}-\imath \frac{a}{3}-\frac{\phi}{3}\Bigg)^{1/2}, \label{chi4}
\end{eqnarray}
\end{widetext}
where the parameters $\gamma$, $\phi$ and $\Delta$ are defined as follows:
\begin{widetext}
\begin{eqnarray}
&&\gamma=\Big(\frac{\phi}{3}-\imath \frac{a}{6}+\frac{12 \pi \sqrt{2/a} A}{3\, \phi}\Big)^{1/2}, \hspace{1em} \phi=\Big(3^{3/2} \Delta-\imath \,a^3 -9\, \imath \pi \sqrt{2 a}A\Big)^{1/3}, \nonumber \\ &&\Delta = \left(26 \pi^2 a A^2-2^{3/2} \pi a^{7/2} A-128 \pi^3\sqrt{2/a^3} A^3\right)^{1/2}. \nonumber
\end{eqnarray}
\end{widetext}

\begin{acknowledgments}
 This work is based upon research supported by the South African Research Chairs Initiative of the Department of Science and Technology and National Research Foundation. The authors thankfully acknowledge the anonymous referee for the useful suggestions.
 \end{acknowledgments}

\end{document}